 \documentclass[smallabstract,smallcaptions]{dccpaper}

\usepackage{epsfig}
\usepackage{citesort}
\usepackage{amsmath}
\usepackage{amssymb}
\usepackage{color}
\usepackage{url}
\usepackage{pdfpages}
\usepackage{cite}

\newlength{\figurewidth}
\newlength{\smallfigurewidth}

\begin{document}

\title
{\large
\textbf{Cool-Chic: Perceptually Tuned Low Complexity Overfitted Image Coder}
}

\author{%
Théo Ladune, Pierrick Philippe, Gordon Clare, Félix Henry and Thomas Leguay\\[0.5em]
{\small\begin{minipage}{\linewidth}\begin{center}
\begin{tabular}{ccc}
Orange Innovation \\
Cesson Sevigne, France\\
\url{firstname.lastname@orange.com}
\end{tabular}
\end{center}\end{minipage}}
}

\Support{Submitted to DCC 2024 in the course of the Challenge on Learned Image Compression (CLIC)\\
© 2024 IEEE.  Personal use of this material is permitted.  Permission from IEEE must be obtained for all other uses, in any current or future media, including reprinting/republishing this material for advertising or promotional purposes, creating new collective works, for resale or redistribution to servers or lists, or reuse of any copyrighted component of this work in other works.}

\maketitle
\thispagestyle{empty}

\begin{abstract}

This paper summarises the design of the Cool-Chic candidate for the Challenge on
Learned Image Compression.
This candidate attempts to demonstrate that neural coding methods can lead to
low complexity and lightweight image decoders while still offering competitive
performance.
The approach is based on the already published overfitted lightweight neural
networks Cool-Chic, further adapted to the human subjective viewing targeted in
this challenge.

The source code is available at \url{https://github.com/Orange-OpenSource/Cool-Chic}
\end{abstract}

\Section{Introduction}


For image coding, most standards are based on conventional designs built upon
the prediction-transformation paradigm. JPEG~\cite{Penn92}, HEVC~\cite{6316136,
7532321} and VVC~\cite{9503377} have successively refined this paradigm through
improved signal processing methods.
\newline

Modern conventional encoders assess the different coding options available and
select the few most suited ones. Encoding is framed as a discrete optimization
problem, i.e. a competition between all parameters of all tools, selecting the
best ones through a rate-distortion (RD) cost.

The compressed signal is sent to the decoder which simply
applies the selected tools to reconstruct the image. As such, the decoder
complexity remains low.
\newline

Successive conventional codecs have provided an ever-increasing
number of hand-crafted coding tools while maintaining a relatively low individual tool
decoding complexity. This leads to an excellent adaptation to the signal and
improved compression performance.
\newline

On the other hand, learned codecs based on auto-encoders~\cite{balle,
elic,jiang2023mlic} offer an alternative coding paradigm. They are designed
to use an offline training stage to optimize a RD cost averaged on a training
set. After training, auto-encoders compute a compressed representation in a
single shot, with no particular optimization of the current image RD cost.
\newline

The current JPEG-AI standardization effort~\cite{jpeg-ai} demonstrates that
auto-encoders are able to outperform conventional codecs~\cite{jpeg-ai-cfp}.
Also, {MLIC++} has recently reported significant improvements over the latest VVC
standard~\cite{jiang2023mlic}. However, the decoder computation effort requires
a significant number of Multiplication Accumulation operations (MAC): between 10,000 MAC
and  1 million MAC~\cite{jpeg-ai-complexity} . This number of MAC implies the
usage of dedicated hardware acceleration (GPUs or TPUs) which deviates from the
hardware conventionally used for image and video decoding schemes.
\newline

Cool-Chic by Ladune et al.~\cite{Ladune_2023_ICCV, Ladune_2023_MMSP} extends the
Implicit Neural Representations \cite{coin} to produce an overfitted image codec
without relying on an autoencoder. Here the image compression process is turned
into an overfitting process where a dedicated lightweight neural network decoder
and latent representation are learned for each image in order to minimize the
individual RD cost.

In this approach the decoder is adapted to the image in conjunction with the
latent domain. The overfitted network parameters and adapted latent variables
are conveyed in a bitstream to the receiver to reconstruct the image.
The difference between the \emph{compressed signal} and the \emph{decoder} is blurred as both the latent variables and the network parameters are representative of the signal to reconstruct.
\newline

By reintroducing an instance-wise RD optimization, this overfitting-based
approach provides compelling coding performance with a low decoder complexity.
Overfitting allows automatic learning of the adapted decoder for each image and
rate constraint.
\newline

This yields compression performance slightly better than
HEVC~\cite{Ladune_2023_MMSP} and with a decoder complexity close to the
conventional decoders such as HEVC or VVC.
\newline

This paper offers a candidate for the Challenge on Learned Image Compression
based on Cool-Chic and tuned inline with the challenge objectives. The objective
is to assess the quality of a perceptually adapted Cool-Chic compared to other
technologies.
\newline

The following aspects are covered:
\begin{enumerate}
    \item Adaptation of the Cool-Chic distortion metric toward the perceptual objective;
    \item Bitstream selection based on a perceptually driven metric;
    \item A section reports the results obtained compared to other codecs.
\end{enumerate}

\Section{Brief description of Cool-Chic}

Cool-Chic is based on the joint training of a decoder and a latent
representation. As such, it does not have an actual encoder as the training
process effectively derives both the decoder and the coded representation.

\SubSection{The Cool-Chic decoder}


The latent representation, which can be assimilated to a frequency domain
representation, bears resemblance to a wavelet decomposition. They are organised in a
pyramidal fashion covering 7 resolutions ranging from the image resolution to
$\frac{1}{64}$ of the image width and height. The latents are synthesised
through the synthesis transform to recover the decoded image. The first step
implies an upsampling stage, organised in a cascade of a $\times 2$ upsamplers. The
upsampling kernel of size $8 \times 8$ is trained. The upsampling stage yields 7 full
resolution latents at the initial image resolution.

These full resolution latents are processed by a succession of 4 simple layers:


\begin{itemize}
\item The first layer is a multilayer perceptron (MLP) with the 7 latents as inputs, and 40 outputs, a ReLU (REctified Linear Unit) non-linearity follows;
\item the second layer is an MLP with 40 inputs and 3 outputs, with a ReLU;
\item the third layer is a residual convolution layer with 3 inputs and 3 outputs, the convolution kernel is 3x3, ReLU operator follows;
\item the last layer is also a 3x3 residual convolution layer with 3 inputs and delivers the R,G and B channels of the reconstructed image.
\end{itemize}

The building elements are voluntarily small in order to limit the complexity:
the reconstruction process involves approximately 600 operations per pixel.
\newline

In order to convey the pyramidal latent representation, an auto-regressive
entropy decoder is used, similar to the Loco-I algorithm~\cite{weinberger1996loco} part
of the JPEG lossless design. To decode a latent value, a context is derived from
the previously decoded values. From that context a mean and a variance are
computed with a 3-layer neural network. Then, the mean and variance drive an
arithmetic range decoder to retrieve the latent values from the bitstream.
\newline

The context decoder neural network has the following components:
\begin{itemize}
\item The first layer is an MLP fed with the 24 closest causal latents  with 24 outputs, followed by a ReLU;
\item the second layer is an MLP with 24 inputs and 24 outputs, also followed by a ReLU;
\item the third and last layer is an MLP with 24 inputs that provides the mean and variance that strives the range decoder;
\end{itemize}

The arithmetic coding is conducted independently on each latent channel enabling them to be run in parallel.
\newline

As the causal context uses 24 neighbouring samples, the complexity of the
entropy neural network decoder is in the range of 1200 MAC. Provided that the
latent representation has 1.33 values per decoded pixel, the complexity of the
arithmetic decoder is 1600 MAC per image pixel.
\newline

Overall the whole decoding architecture has a maximum complexity of
approximately 2200 MAC. In practice many latents values can be zeroed out by the
optimization process, lowering the complexity especially at lower rates.
\newline

The computing elements in Cool-Chic are also limited. This is directly reflected
in the decoder size which is in the order of a few dozen of kilobytes.

\SubSection{Encoding and training loss}

The encoding process is a simple training process: the pyramidal latents,
entropy neural network decoder and latent synthesis modules are initialised with random
values. Then a standard gradient optimisation is conducted to tune the different
neural network parameters and latent values.
\newline

The loss used for this training process is based on a Lagrangian multiplier that
balances the trade-off between the rate and the distortion.

\begin{equation}
J(\lambda) = D(x,\hat{x}) + \lambda \cdot R(\hat{y})
\end{equation}

where $\hat{x}$ is the decoded image obtained using the latents $\hat{y}$,
through the decoder $\psi_{\theta}$ parameterised by the neural network
parameters~$\theta$.

\begin{equation}
\hat{x} = \psi_{\theta} (\hat{y})
\end{equation}

In order to take into account the perceptual objective of this challenge, the
distortion used in this paper combines the \emph{mse} (mean squared error) and the
MS-SSIM metric~\cite{1292216}.

\begin{equation}
D(x,\hat{x}) = || x - \hat{x} ||^2 + \alpha \cdot (1-\text{msssim}(x , \hat{x} ))
\end{equation}

The factor $ \alpha=0.01 $ is chosen empirically to balance the \emph{mse} and MS-SSIM.

During the encoding process a selection of different $\lambda$ values is
provided to obtain a range of bit rates compatible with the challenge objective.

\Section{Coding results}

\begin{figure}[t]
\begin{center}
   \includegraphics[width=0.8\linewidth]{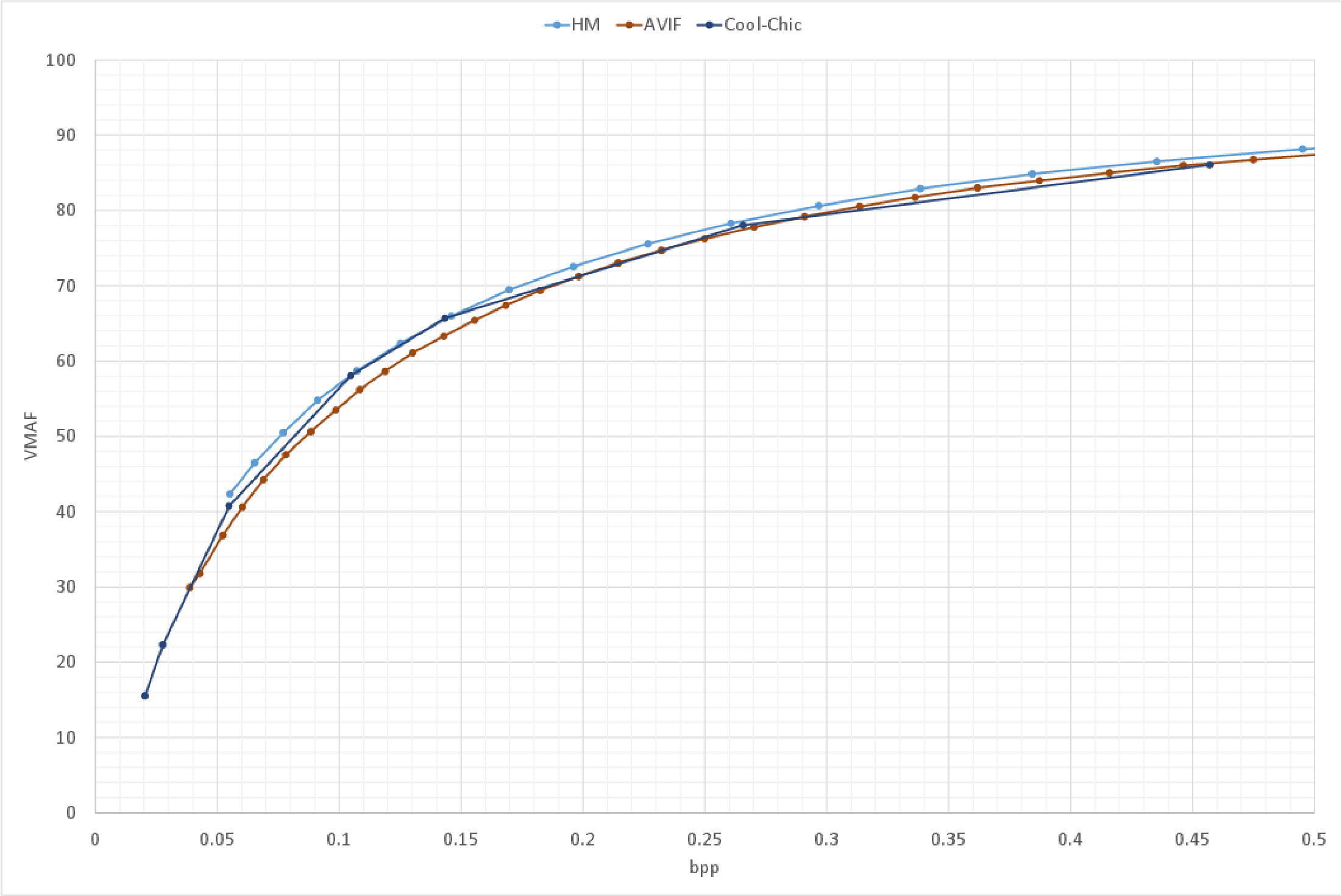}
\end{center}
   \caption{Comparison of the VMAF scores obtained by the Cool-Chic submission relative to AVIF (optimised with --tune ssim) and the HM.}
\label{fig:coding}
\end{figure}

Figure~\ref{fig:coding} draws the characteristics for the HEVC test model (HM),
AVIF (the AV1 coder for images), tuned with ssim and the results of Cool-Chic
for 6 selected $\lambda$ values.

The VMAF~\cite{vmaf} metric is used as it is a good indicator of the image quality.
\newline

The level of performance of the three codecs seems in the same range: through
the perceptual optimisation of Cool-Chic, this codec is in the same performance
range of AVIF and HM.
\newline

For the challenge 0.075, 0.150 and 0.300 bit per pixels on average are defined
for the image coding challenge. For the validation set, the average picture
resolution is around 3 million pixels, and the total allowed byte budget is
respectively 823660, 1647321 and 3294643 bytes for the three bpp targets.
\newline

For this candidate the individual bit budget per image is selected based on the
VMAF~\cite{vmaf} metric. For each target bpp, the coding level for each sequence
is chosen to meet the bit budget while maximising the minimum VMAF among the
images.
\newline

Table~\ref{tab:metrics} summarises some objective measures for this submission.
The PSNR and MS-SSIM are the ones reported on the validation leaderboard. The
VMAF metrics are also reported, through the average over the sequences and the
lowest VMAF among the 30 validation pictures (which is the criterion maximised
here).
\newline

\begin{table*}
\centering
\begin{tabular}{l c c c c c c}
\hline
	 bpp &   {Data size} 	&   average VMAF 	&  worse VMAF 	&    PSNR 	&    MS-SSIM	\\
	0.075 &   823530 		&   48.305       	&  46.128 		&    26.533 &    0.927 	\\
	0.150 &   1646858 		&   65.482 			&  63.861 		&    28.925 &    0.958 	\\
	0.300 &   3293682 		&   79.005 			&  77.577 		&    31.686 &    0.976 	\\
\hline
\end{tabular}

\caption{Objective metrics for the 3 challenge bit rates for the candidate}
\label{tab:metrics}
\end{table*}

\begin{figure}[t]
\begin{center}
   \includegraphics[width=0.8\linewidth]{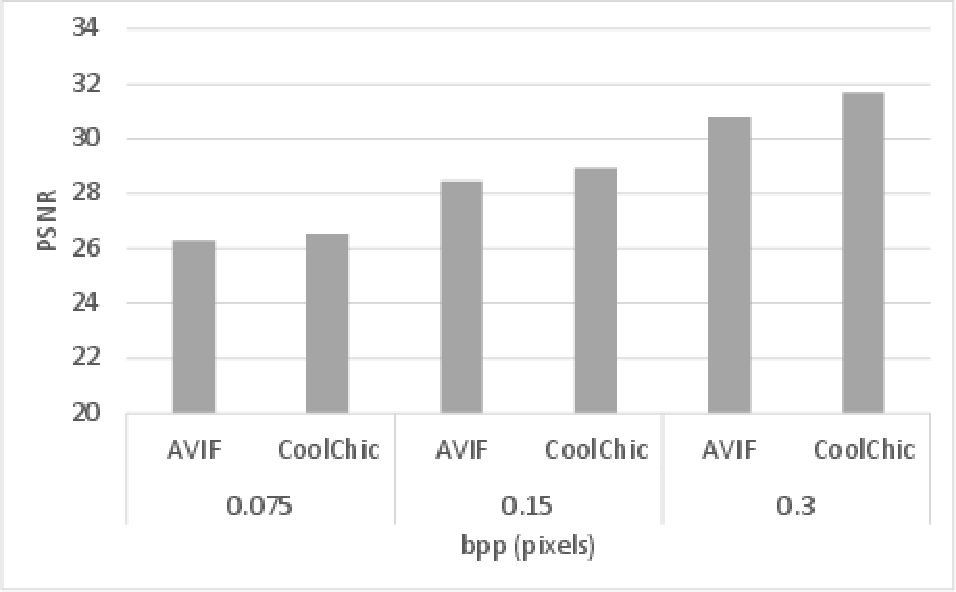}
\end{center}
   \caption{Comparison of the PSNR scores obtained by the Cool-Chic submission relative to AVIF (optimised with --tune ssim).}
\label{fig:psnr_avif}
\end{figure}

\begin{figure}[t]
\begin{center}
   \includegraphics[width=0.8\linewidth]{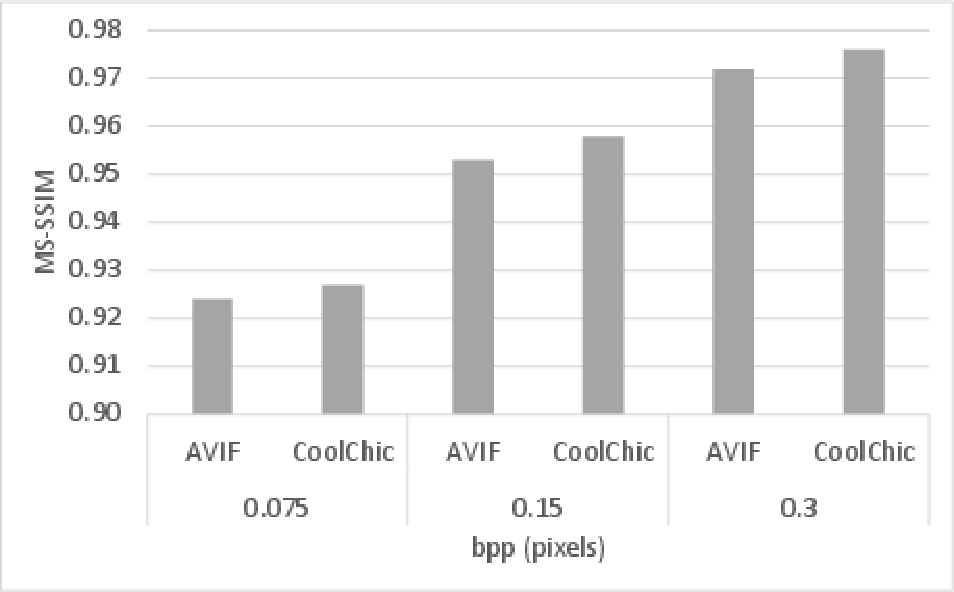}
\end{center}
   \caption{Comparison of the VMAF scores obtained by the Cool-Chic submission relative to AVIF (optimised with --tune ssim).}
\label{fig:msssim_avif}
\end{figure}

As displayed in figures~\ref{fig:psnr_avif} and~\ref{fig:msssim_avif}, the PSNR
and MS-SSIM metrics for Cool-Chic are close to those of AVIF and consistently
above for the three bit rates.
\newline

This confirms that this lightweight neural network codec is competitive
with AVIF.

\Section{Conclusion}

This paper provides a perceptually tuned version of Cool-Chic. To provide a set
of images targeting subjective evaluation, Cool-Chic encoding is simply adapted
through the use of a weighted distortion metric.

The objective measurement through the VMAF metric seams to indicate that,
despite its simple structural complexity, the quality level of Cool-Chic is
comparable to AVIF and HEVC.

\Section{References}

\bibliographystyle{IEEEbib}
\bibliography{refs}

\end{document}